\documentclass[12pt]{article}
\usepackage{natbib}
\begin{document}
\setcitestyle{semicolon,aysep={}}
\date{}
\title{Identical quantum particles as distinguishable objects}
\author{Dennis Dieks and Andrea Lubberdink \\History and Philosophy of Science\\
Utrecht University}
\maketitle
\begin{abstract}
According to classical physics \textit{particles} are basic building blocks of the world. Classical particles are distinguishable objects, individuated by physical characteristics. By contrast, in quantum mechanics the standard view is that particles of the same kind (``identical particles'') are in all circumstances  indistinguishable from each other. This indistinguishability doctrine is motivated by the (anti)symmetrization postulates together with the assumption (``factorism'') that each single particle is represented in exactly one factor space in the tensor product Hilbert space of a many-particles system. 

Although the factorist assumption is standard in the literature, it is conceptually problematic. Particle indistinguishability is incompatible with the everyday meaning of ``particle'', and also with how this term is used both in classical physics and in the experimental  practice of physics. The standard view requires quantum particles of the same kind to remain indistinguishable even in the classical limit, so that a transition to the classical picture of the world seems impossible.

In \citep{lubberdink,lubberdinkb,die5} we have proposed an alternative conception of quantum particles that does not rely on factorism and that avoids these and other problems. We further explain and develop this proposal here. In our view, particles in quantum theory are not fundamental but \emph{emergent}. However, in those situations where the particle concept \emph{is} applicable, quantum particles---identical or not---are distinguishable in the same way as classical particles.    
 \end{abstract}

\section{Introduction}

The physical world of everyday experience consists of individual objects that can be distinguished from each other by distinctive physical characteristics. Classical physics generalizes this picture into the microscopic domain by the introduction of the notion of a \textit{particle}: a classical particle is characterized by an individuating set of values of physical quantities (mass, electric charge, position in Newtonian absolute space, momentum, etc.). Although different classical particles may have some of these values in common, there never is a complete overlap because according to classical theory repulsive forces become increasingly strong (and eventually diverging) when mutual particle distances decrease, so that any two particles differ at least in their spatial locations.

We can associate an individual name or a label with each classical particle: $1,2,...,N$ for a collection of $N$ particles. This labeling can take the form of providing explicit definite descriptions that identify particles by specifying their unique characteristics (e.g., by statements like ``Call the particle in the bottom left corner of the container particle $1$, the particle one centimeter above it particle $2$, etc.''). Often, however, it is taken for granted that distinguishing characteristics exist without actually using them to fix the labels. In this case the reference of the labels is left open as a matter of convention, and arbitrary permutations of the labels are not considered to  correspond to a change in the physical situation that is described. 

In all cases, an assumption tacitly made in physics is that there is no particle identity beyond what can be based upon physical characteristics (values of physical quantities including their time evolution, trajectories, and if needed relational properties, see section \ref{wd})---this means a rejection of metaphysical principles like haecceity or primitive thisness. 

\textit{Which} physical quantities are relevant for describing particles of a particular kind is governed by law-like physical principles \citep{lombardidieks2}. A distinction must been made between two sorts of such quantities: the quantities that take values that vary depending on the state of the particle (e.g., position, velocity), and the quantities that are state independent and typical of the kind of particle we are dealing with (e.g., electron rest mass, electron charge).

Quantum mechanics also describes elementary quantum systems, e.g.\ electrons, by means of state dependent and state independent quantities. There is an important difference with classical mechanics, however: the state dependent quantities are now represented by (not necessarily commuting) \textit{operators} instead of numerical values, and instead of the classical state (or ``phase'') spaces we have to use vector spaces (Hilbert spaces). Nevertheless there are strong similarities between the structures of quantum and classical mechanics: for example, many functional relations between classical quantities are taken over in quantum mechanics as relations between operators. Where Poisson brackets between physical quantities occur in the formulas of classical theory, commutators between operators appear in quantum formulas. 

It is important for the subject of this paper to note that there is also an analogy between classical mechanics and quantum mechanics with regard to how many-particles state spaces are built up from one-particle state spaces. Both classically and quantum mechanically, an $N$-particles state space must be able to accommodate all kinematically possible combinations of one-particle states. In classical physics, the state of one particle of a given kind is fixed by the values of a certain number, $n$ say, of (state dependent) physical quantities. Therefore we need the specification of $n \cdot N$ quantities to determine the total state of $N$ particle of the same kind; the $N$-particles state space must consequently be an $n \cdot N$-dimensional manifold of points (each point representing a possible state). If the one-particle state spaces are discrete, with $X$ possible states in each of them, the number of states in the $N$-particles state space in this way becomes $X^N$. Similarly, in quantum mechanics an $N$-particles Hilbert space must be able to accommodate all possible combinations of one-particle states. However, the one-particle states are now not defined by the specification of the values of a fixed set of numerical quantities, as in classical mechanics. Instead they are  defined as linear superpositions of independent basis states (basis vectors) in a one-particle Hilbert space; let us assume, for the sake of illustration, that there is a finite number $X$ of such independent basis states (i.e., the one-particle Hilbert space is $X$-dimensional). In this case the combination of all  basis vectors of $N$ one-particle spaces leads to a set of $X^N$ new independent basis states, spanning the $X^N$- dimensional ``tensor'' product space ${\cal{H}}^N = {\cal{H}}_1 \otimes {\cal{H}}_2 \otimes {\cal{H}}_3 \otimes ... \otimes {\cal{H}}_N$.  

In this expression for the total state space (the tensor product), the one-particle Hilbert spaces (the factor spaces) are labeled, $1,2,...,N$. This is analogous to the labeling of individual particle spaces, and the quantities defined in them, in the classical case. A point in an $N$-particles classical state space (representing $N$ particles of the same kind) has coordinates $(Z_1, Z_2, Z_3, ... , Z_N)$, where $Z_i$ denotes the set of quantities defining the state of particle $i$ (which is different from the states of all other classical particles, as noted before). Because each classical one-particle space represents one particle with its unique state, different from the states of the other particles, any labeling used for the one-particle spaces can also be used to label the particles themselves (and \textit{vice versa}). Usually these labels are of the second category mentioned above: the numbering is introduced abstractly, without fixing them by explicit reference to specific properties, with the consequence that the labels may be permuted arbitrarily without changing anything in the physical situation that is described. By contrast, in discussions of concrete cases the labels of both state spaces and single particles are often explicitly defined via definite descriptions, so that they are fixed and can no longer be permuted. In this case the labels are shorthand for the individuating one-particle states (e.g., the label $1$ refers by definition to the particle that at time $t=0$ found itself at $x=0$, and so on). As said before, in classical mechanics this definite labeling via individuating states is  always possible in principle.    

In view of the analogies between how the classical and quantum many-particles state spaces are constructed it seems plausible to assume that also in the quantum case one-particle factor state spaces (Hilbert spaces) and their labels are in a one-to-one relation with single particles and their states in an $N$-particles system. This is in fact the standard textbook view: factor space labels are associated with single particles, which allows us to speak of particle $1$, particle $2$, and so forth.\footnote{This standard doctrine---that factor space labels refer to single particles---has been baptized ``factorism'' by Caulton \citep{caulton,leegwater}.\label{factorismfootnote}}  

However, as we shall argue, there is in fact \textit{no} valid parallelism between classical and quantum mechanics  on this point: the factor space labels in the quantum mechanics of particles of the same kind should \emph{not} be thought of as referring to single particles. Our argument will crucially depend on a specific feature of the quantum formalism, namely that although the $N$-particles \textit{state space} possesses the structure of an $N$-fold product of Hilbert spaces, the many-particles \textit{states} in this space \textit{do not} have the form of a concatenation of one-particle states, each in its own factor space. This complication is due to the symmetrization postulates, which prescribe that  $N$-particles states should be symmetrically entangled superpositions.

\section{The symmetrization postulates}\label{symmetrization}

Consider a product state of the form $| L \rangle_1 |R \rangle_2 $\footnote{We consider this state for the sake of argument: as will become clear in a moment, such product states of particles of the same kind are not allowed in quantum mechanics.}, defined in a two-particles tensor product Hilbert space for particles of the same kind ${\cal{H}}_1 \otimes {\cal{H}}_2 $, with $| L \rangle $ and $|R \rangle $ standing for wave functions with support in narrow left and right regions of space, respectively.\footnote{The labels $1$ and $2$ are not strictly necessary here and in similar expressions: the factor spaces can be identified by their order, from left to right, in the expressions. For ease of verbal reference we will go on using them, though.} The natural interpretation is that this state represents one particle located on the left, and one on the right. Indeed, appropriate position measurements will with certainty result in one successful particle detection in the $L$ region and one in the $R$ region. The  particles in question are individuated by the orthogonal states $| L \rangle$ and $|R \rangle $, and one may consequently introduce particle labels in the same way as in classical physics. In fact, $L$ and $R$ themselves can function as particle labels. 

Now reverse the order of the two particle states, so that the state $| R \rangle_1 |L \rangle_2 $ results. The particle that was originally represented in factor space $1$, namely particle $L$ (the left one) is then represented in factor space $2$, and the other particle is described in factor space $1$. This brings out a conceptual difference, in principle, between particle and factor space labels---a point that will be significant in our general analysis of the meaning of particle labels. However, in our present example this complication appears to be without importance. Indeed, given that particles are individuated solely by  physical quantities (in our case only by $L$ and $R$), the two states $| L \rangle_1 |R \rangle_2 $ and $| R \rangle_1 |L \rangle_2 $ represent the same physical situation, so that we are facing a case of theoretical surplus structure. It is natural to remove this superfluous multiplicity of descriptions by defining one single representation: stipulate, by convention, that we call factor space $1$ the one representing particle $L$, and similar for the particle $R$. We thus obtain the unique representation  $| L \rangle_1 |R \rangle_2 $. 

This would be the same procedure as the one followed in classical mechanics. As mentioned in the Introduction, there is a conventional choice to be made when we label classical particles, and this can be used to stipulate that particle $i$ is the one whose state is an element of the factor space labeled $i$\footnote{This is all on the assumption that the particles are of the same kind. When two particles belong to different kinds, they are individuated by state independent physical properties that also distinguish the factor spaces. In that case swapping of states as discussed here will not be possible.}. The multiplicity of theoretical descriptions in classical mechanics is thus dealt with as a harmless consequence of the conventionality of labeling. We may conventionally select one set of labels and work with the single resulting description\footnote{A rarely seen alternative is to take the full collection of states that relate to each other by label permutations as together representing the physical situation---this ``super-state'' has been called the Ehrenfest Z-star \citep{ehrenfest,die5}. Another, more popular way of removing the multiplicity in classical mechanics is to go over to the \textit{reduced phase space}, by identifying phase points that are mapped to each other by permutations of the labels.}. 

It is a very remarkable feature of quantum mechanics, however, that the multiplicity of physically equivalent states that we just discussed cannot occur at all. It is decreed, by quantum law, that for systems of particles of the same kind product states like $| L \rangle_1 |R \rangle_2 $ and $| R \rangle_1 |L \rangle_2 $ are \textit{forbidden}. Only a third type of state, namely an (anti)symmetric superposition of these product states, is allowed.      

More precisely, there are two law-like postulates: a symmetrization postulate for bosonic systems, permitting only states that are invariant under permutations of the labels, and an antisymmetrization postulate for fermions that introduces a minus sign for uneven permutations. Thus, instead of the two physically equivalent product states of our example only one of the following states is allowed:
\begin{equation}\label{symm}
 |\Psi \rangle = \frac{1}{\sqrt{2}} \{ | L \rangle_1 |R \rangle_2 \pm  | R \rangle_1 | L \rangle_2 \},
\end{equation}
where the plus sign holds for the bosonic case and the minus sign for fermions.

The states $|L\rangle$ and $|R\rangle$ in Eq.(\ref{symm}) therefore occur symmetrically in the factor spaces labeled by $1$ and $2$, respectively. As a consequence, if in analogy with the classical case we assume that these labels not only refer to the factor spaces but also to the single particles composing the total system, we have to conclude that all these particles are in exactly the same state. Indeed, all factor spaces contain exactly the same states in exactly the same way so that the properties of the particles described in these factor spaces, as given by standard quantum mechanics, are the same. it therefore becomes impossible to individuate them by their physical characteristics.

In a more formal way this conclusion can be reached by determining states associated with the individual factor spaces via the procedure of taking ``partial traces'': tracing out over the parts of the total state labeled by $2$ we obtain the mixed state $ W = 1/2 \{| L \rangle \langle L | + | R \rangle \langle R | \} $; and exactly the same state by tracing out over $1$. These two identical mixed states represent the one-particle quantum states as defined in the single factor spaces $1$ and $2$.
This conclusion generalizes to (anti)symmetric $N$-particle states, with $N>2$: each factor space label is associated with the very same mixed state.

\section{Problems of factorism}

The correlation between one-particle state spaces on the one hand and individuating particle states (one unique state for each particle) on the other, which in classical mechanics justifies the double use of labels as both referring to one-particle phase spaces and to uniquely identified particles thus breaks down in quantum mechanics.  
If quantum particles correspond to factor space labels in the way factorism wants it,   they must all be in exactly the same one-particle state and therefore they must all possess exactly the same physical properties. Although there have been discussions of this feature, in particular in connection with the notion of ``weak discernibility'' (more about this in section \ref{wd}), we believe that in the literature it has not been appreciated sufficiently how weird, and adverse to the very idea of a particle, this consequence of factorism actually is. As the symmetrization postulates apply to the collection of all particles of any given kind in the whole universe, for example all electrons, the factorist must hold that each single electron is equally present at all positions in the universe at which there is ``electron presence''. So according to factorism it should not make sense to speak about the specific electrons in my body, e.g., since all electrons in the universe are equally present there.  This result is not restricted to localization but holds in the same way for whatever physical particle property one may think of. All electrons, and in  the same way all protons, neutrons and all other particles of the same kind are mutually indistinguishable.

This leads to a  lack of individuality that is in conflict with the very notion of a particle, and is in stark contrast not only to how the notion of a particle is used in classical physics but also to how it is used in physical practice. This is important, because the very motivation for speaking about particles also in the context of quantum mechanics derives from analogies with classical physics and from the use that can be made of the particle concept in experimental practice.

To see a concrete illustration of the problem, consider the Einstein-Podolsky-Rosen-Bohm state. In the foundational literature the EPR thought experiment (in its modern spin version due to Bohm) is standardly discussed as pertaining to two electrons at a large distance from each other, with a total spin state $\frac{1}{\sqrt{2}} \{ |\!\uparrow \rangle_1 |\downarrow \rangle_2 - |\!\downarrow \rangle_1 |\uparrow \rangle_2 \}$. The spatial part of the wave function is often not written down explicitly, but it must be considered as well in order to make contact with the locality question that is at issue in the EPR experiment. The total state including this spatial part has the form
\begin{equation}\label{EPReq}
| \Phi \rangle = \frac{1}{\sqrt{2}} \{| L \rangle_1 | R
\rangle_2 + | R \rangle_1 | L \rangle_2 \} \otimes \{ |\!\uparrow
\rangle_1 |\downarrow \rangle_2 - |\!\downarrow \rangle_1
|\uparrow \rangle_2 \},
\end{equation}
where $| L \rangle$ and $| R \rangle$ as before are states localized on the left and right, respectively, at a large distance from each other. In the language of wave mechanics, $| L \rangle$ and $| R \rangle$ represent narrow wave packets.
Note that the spatial part of $ | \Phi \rangle $ is symmetric in the labels $1$ and $2$, whereas the total state is antisymmetric as required by the antisymmetrization postulate (we are dealing with fermions).

Now, if we are to accept the factorist position that the labels $1$ and $2$ in Eq.(\ref{EPReq}) refer to the two EPR particles, we have to reconcile ourselves to the idea that there is neither a left nor a right electron. The spatial states associated with both $1$ and $2$ are exactly the same, namely $ 1/2 (| L \rangle \langle L | + | R \rangle \langle R | ) $, so that each of the corresponding particles would be ``evenly spread out'' over left and right. This means that the way the EPR case is standardly understood in foundational discussions and in experimental practice, as being about two (more or less) localized systems at a large distance from each other, is at odds with the official theoretical account, namely the factorist interpretation of the indices $1$ and $2$ as particle labels\footnote{As we shall discuss in sections 5 and 6, the rejection of factorism makes it possible to interpret the state of Eq.(\ref{EPReq}) as a representation of two localized systems. However, as we shall see, the particular (``non-trivial'') form of the superposition in (\ref{EPReq}) stands in the way of a full localized particle interpretation of this state. This is behind the non-locality of state (\ref{EPReq}) that is manifested by violations of Bell inequalities.}. 

The problem is aggravated by the observation that the strange features of factorist particles persist in the classical limit of quantum mechanics. The symmetrization postulates are meant to be generally and universally valid, in all physical situations; they are not affected by limits and approximations, whatever the exact details of these limiting and approximation procedures may be. This implies that the sameness of partial traces in all factor spaces is a general and robust feature of quantum mechanics that survives in the classical limit, with the consequence that even in this limit all quantum particles possess exactly the same properties. Of course, this sameness of properties cannot actually be upheld for the particles that occur in the classical theory that results from a successful limiting procedure. So if factorism were right, the particles that we know from classical physics could not correspond to their quantum namesakes---e.g., in the limit there would be no transition from electrons in quantum theory to electrons in classical electrodynamics. That is a strange and undesirable predicament: as pointed out before, the very introduction of the particle concept in physics is motivated by classical experience and we expect at the very least that the quantum concept approximates the traditional one in classically describable situations.   

Before explaining our alternative quantum particle notion that avoids these problems, we want to discuss the attempt to salvage factorism  by dropping the requirement  that particles be distinguishable by their own individual (monadic) characteristics. The idea of this proposal \citep{saundersleibn,saundersobj} is that it is sufficient for objects of the same sort in general, and identical quantum particles in particular, to be provably different from each other by virtue of the relations they stand in, even if it is impossible to refer to them individually by physical means.

\section{Weak discernibility}\label{wd}

According to classical physics it is theoretically possible to have objects that are different from each other even though they possess exactly the same properties. A famous example was proposed and discussed by \citet{black}\footnote{\citet{leegwater} list and discuss also several other, similar examples from the philosophical literature.}: consider two perfect spheres of exactly the same material constitution, alone in relational space (in order to exclude absolute position as a distinguishing property), at a mutual distance of $1.5$ miles. By stipulation, no physical features are able to distinguish these two spheres; but still there are \textit{two} of them. This seems a consistent theoretical possibility. Consequently, it appears that the individuality of these spheres cannot be reducible to physical differences (which seems to signal a violation of Leibniz's Principle of the Identity of Indiscernibles). 

However, as pointed out by \citet{saundersleibn,saundersobj}, who takes his cue from \citet{quine}, Black's spheres stand in an \textit{irreflexive} physical relation to one another: a relation that an entity cannot have with respect to itself. Indeed, each sphere has a non-vanishing distance to a congruent sphere of identical composition. The irreflexivity of this relation (a sphere cannot have a non-zero distance to itself, assuming the usual Euclidean topology) makes it possible to satisfy a form of Leibniz's Principle after all: if a sphere stands in a physical relation that it cannot have to itself, it logically follows that there must be at least two spheres. The spheres' numerical diversity can thus be grounded physically. Of course, in cases like this it remains impossible to identify the objects by means of physical descriptions, since any description applicable to one object applies equally well to all others---it is only the numerical diversity that can be physically grounded\footnote{As \citet{friebe} notes, Leibniz's Principle is therefore not saved in the sense that it bestows a physically underpinned individuality on each object.}. Objects like Black's spheres (equal in all respects but still numerically diverse by virtue of irreflexive relations) are called ``weakly discernible''.

\citet{saundersleibn,saundersobj} has suggested that identical quantum particles defined in the factorist way are like Black's spheres. The simplest case is that of identical factorist fermions: although the states in the different factor spaces are the same, irreflexive relations may be argued to exist between parts of the total state associated with different labels. For example, in the singlet state $\frac{1}{\sqrt{2}} \{ |\!\uparrow \rangle_1 |\downarrow \rangle_2 - |\!\downarrow \rangle_1 |\uparrow \rangle_2 \}$, the relevant relation is ``having opposite spin directions''. 
From this \citet{saundersleibn,saundersobj} and \citet{muller1} conclude that the objecthood of factorist fermions can be physically grounded. \citet{muller2} extend this argument to bosons. They argue that quite generally physical differences are associated with different  factor spaces: operators (representing physical quantities) that belong to different factor spaces always commute, whereas this need not be the case for operators defined within one and the same factor space. In particular, momentum and position operators with different factor labels always commute, but do not when defined in the same factor space. This fact can be exploited to define irreflexive relations, which in turn can be used to argue that even bosons are weakly discernible objects.

There are several controversial points here, though (see for a more extensive account \citep{die4,die5}). First, any argumentation for weak discernibility can only have force if the irreflexive relations that are invoked are physically relevant. In the present context this issue boils down to the question of whether relations between mathematical quantities defined in different factor spaces may be interpreted as representing relations between (candidate) physical objects---however, this is precisely the issue under discussion in the debate about factorism. Put differently, the irreflexivity argument by itself does not show that factor labels refer to physical particles: the irreflexive relations between different factor spaces can only be used to argue for weak discernibility of physical objects if it already is assumed that the factor spaces correspond to such objects. {If} factorism is assumed to be right, \textit{then} we may conclude that the particles represented in the different factor spaces are  numerically different on physical grounds. Without the presupposition of factorism we would just be studying mathematically defined irreflexive relations between component spaces of the total Hilbert space.  

A second point to note is that the relations considered in the weak discernibility arguments are represented by Hermitean \textit{operators}, whose interpretation as physical \emph{properties} can be disputed. Indeed, the standard quantum doctrine is that such operators are to be used for calculating measurement results, and that the outcomes of  measurement interventions should not be mistaken for what there was \textit{before} the measurement was performed. So although it is true that double spin measurements in the singlet state result in opposite spin values, it is not self-evident that this is translatable into a statement about a pre-existing relation between objects. Indeed, in the context of the EPR-experiment the usual interpretation of the correlations predicted by the singlet state is that these correlations testify to the holistic character of the spin system and should not be explained by pre-existing relations between spin properties of the parts. In fact, as argued in \citep{die4}, such correlations between measurement results can also be understood without assuming a particulate picture. 

But the most important objection from our point of view is that the introduction of weak discernibility does nothing to alleviate the unnaturalness and even weirdness of factorist particles. As noted before, these supposed entities are omnipresent, find themselves all in exactly the same state and share all their physical properties. They are therefore extremely different from what we are used to call particles in ordinary language and classical physics. Their potential weak discernibility, which makes them numerically distinct even though they have all their properties in common,  only adds to their mysteriousness and does not make them more acceptable as particles.  We should surely strive for a quantum particle concept that is less extravagant and closer to the ordinary concept of a particle; one that approximates the classical notion in the classical limit. In order to achieve this we must abandon factorism.

\section{Quantum particles}\label{QP}

An alternative way of defining and identifying quantum particles was proposed in \citep{lubberdink,die5}. Its basic idea is to associate particles not with the labels of the factor spaces, but instead with \textit{one-particle states} that occur in the total $N$-particles state. In fact, this is precisely the motivating thought at the original basis of factorism itself: the viability of the classical version of factorism was warranted by the strict correlation between one-particle state spaces and unique one-particle \emph{states} in classical theory, as discussed in sections 1 and 2. This correlation breaks down in quantum mechanics, which spells the demise of factorism. But the idea of identifying particles by their states and properties certainly remains eminently reasonable.

To see how this alternative notion can be implemented, consider again the antisymmetric state 
\begin{equation}\label{LR}
\frac{1}{\sqrt{2}} \{ | L \rangle_1 |R \rangle_2 -  | R \rangle_1 | L \rangle_2\},
\end{equation}
in which $| L \rangle $ and $ |R \rangle $ stand for two non-overlapping wave packets at a large distance from each other---one packet located on the left, the other on the right. As we have seen, according to factorism this state represents two particles that are both in the state $ W = 1/2 \{| L \rangle \langle L | + | R \rangle \langle R | \} $, so both equally ``smeared out'' over left and right (and---perhaps, see our earlier discussion---weakly discernible because measurements will show an anti-correlation between $L$ and $R$ results of double position measurements). However, the nature of this state, with its two widely separated and narrow spatial regions in which something can be detected at all, makes one rather expect that (\ref{LR}) represents a situation with one particle on the left and one on the right; the results of position measurements would confirm this interpretation. Actually, this interpretation is already silently adopted in the actual practice of physics (for example in  EPR discussions). In order to work out this alternative interpretation the particles should apparently be associated with the states $| L \rangle $ and $|R \rangle $, respectively, even though each of these states occurs in \textit{both} factor spaces. 

In \citep{lubberdink,die5} this idea was proposed and worked out with special attention for the case in which the  one-particle states occurring in the total state do not overlap in three-dimensional space---this avoids issues relating to the non-uniqueness of the decomposition of states like (\ref{LR}) and forges a bridge to the classical picture, in which particles are always localized. The essential idea, however, is more general: it is to associate particles with orthogonal one-particle states instead of factor labels, and this may be possible also in situations where the states do not have a spatial interpretation (cf.\ \citep{caulton}). As it turns out, a general theoretical framework can be used here that was proposed by \citet{ghirardi}.        

\citet{ghirardi} start from the observation that all observables (operators representing physical quantities, which in principle can be measured) of systems of identical quantum particles must be symmetric in the factor labels. That this has to be so can be seen from the fact that asymmetric operators break the symmetries required by the symmetrization postulates when they operate on states, so that they are inconsistent with these postulates. It may also be concluded from the physical consideration that interaction Hamiltonians must be symmetric because factor labels in themselves are not measurable  \citep{die0}.  Candidate individuating particle properties can accordingly be represented by symmetric projection operators (telling us whether or not a property is possessed, via the eigenvalues $1$ and $0$, respectively). That means that we should not be concerned with operators of the form $ P_1 \otimes I_2 $, with $I_2$ the unity operator in factor Hilbert space $2$, if we want to investigate particle properties, but rather with projection operators like
\begin{equation}\label{symproj}
   P_1 \otimes I_2 + I_1 \otimes P_2 - P_1\otimes P_2 ,
\end{equation}
with $P$ standing for the projection operator to be used in the case of a one-particle system (in which there is only one factor space). The expectation value of the operator in (\ref{symproj}) in an (anti)symmetric state yields the probability of finding at least one particle with the property represented by $P$ when we perform a measurement. The last term in (\ref{symproj}) can be left out in the case of fermions.\footnote{The last term of (\ref{symproj}) was added to allow for the possibility that the same one-particle state occurs twice in the total state, which may happen in a bosonic state. Without the last term the probability would become greater than $1$ in this case. In fermion states one-particle states cannot occur more than once. }

As \citet{ghirardi} observe, the use of such symmetric projection operators makes it sometimes possible to associate a set of pure one-particle quantum states with a many-particles system even in the symmetric and antisymmetric total states required by the symmetrization postulates. Indeed, such (anti)symmetric total states may be eigenstates with eigenvalue $1$ of symmetric projection operators like (\ref{symproj}), so that the probability to find the corresponding property in a measurement is $1$. When this is the case, the proposal is to associate the one-particle state on which $P$ projects with one subsystem of the many-particles system. 

This procedure leads to the association of $N$ pure one-particle states with an (anti)symmetric $N$-particle state, if the following condition is satisfied: the total (anti)symmetric state is obtainable from symmetrizing or antisymmetrizing an $N$-fold \textit{product state}. Such (anti)symmetrized product states are eigenvectors of symmetric projection operators of the type (\ref{symproj}), if the individual one-particle projection operators occurring in the expression project on the one-particle states that are the factors in the  (anti)symmetric product. The usefulness of this result in the context of our earlier considerations should be clear: in the case of (anti)symmetrized product states one can define a set of one-particle states that are candidates for defining one-particle subsystems. 

In order to successfully employ this procedure for constructing an interpretation in terms of distinct individual subsystems, the one-particle states that we find should be mutually orthogonal. This orthogonality is guaranteed in the case of fermionic systems; it may, but need not, obtain in the bosonic case. In accordance with common usage and classical physics, as explained earlier, we will  reserve the use of the particle concept for subsystems defined by orthogonal states---the thus defined particles are  distinguishable by their states (they are ``absolutely discernible'', i.e.\ distinguishable by means of monadic physical properties). 

Accordingly, fermionic states that are antisymmetrized product states can always be interpreted in terms of distinct particles. Bosonic systems that are symmetrized product states do not always admit a particle description, because the one-particle states occurring in them need not be mutually orthogonal. In such situations bosons are better described as assemblies of field quanta (in a Fock space occupation number representation---see \citep{die4,die5}).  

This proposal for using the notion of a particle (\citep{lubberdink,die5,caulton}) fits the actual practice of physics. 
The two-fermions state (\ref{LR}) can be used for a quick concrete illustration. This state is an eigenstate of the symmetric projection operators $|L\rangle_1 \langle L |_1 \otimes I_2 + I_1 \otimes |L\rangle_2 \langle L |_2 $ and $|R\rangle_1 \langle R |_1 \otimes I_2 + I_1 \otimes |R \rangle_2 \langle R |_2 $, which leads to the conclusion that we have one particle characterized by $|L \rangle $ and one particle characterized by $ | R \rangle $, respectively. The state (\ref{LR}) thus represents one particle at the left and one at the right, even though the labels $1$ and $2$ are evenly distributed over $L$ and $R$. In the transition to the classical limit this will yield what we expect: classical particles are localized entities, to be approximated by narrow wave packets, following approximately classical trajectories.\footnote{Very narrow quantum wave packets spread out very quickly, and will therefore only be able to follow approximately classical trajectories for a very short time. For the classical limit, and the applicability of the classical description, conditions must be fulfilled that counteract this dispersion of wave packets, like the presence of appropriate decoherence  mechanisms.} 

So what we propose is that the identification of quantum particles should be grounded in the distinctness of  physical properties, represented by one-particle projection operators and their mutually orthogonal eigenstates.\footnote{We should mention the important point that the decomposition in terms of such states as given in (\ref{symm}) is not unique. The equality of the coefficients appearing in front of the terms in the (anti-)symmetric superposition is responsible for a degeneracy, so that infinitely many alternative decompositions, in addition to the one in terms of $| L \rangle $ and $| R \rangle $ are possible. So the set of properties that distinguish the quantum particles is  underdetermined by the procedure as we have outlined it. To make the definition of the particles unique some additional ingredient is needed, which picks out a privileged particle-properties basis. One possibility is to postulate the position basis as privileged  \citep{lubberdink,die5}; this ties in with the localized nature of particles in the classical limit. An alternative and more general idea is to assume that properties and states have to be defined \textit{relative} to an external ```observing'' system, and that the interaction with the external system determines the property basis---cf.\  \citep{die9} for an exploration of this idea. In many cases the latter proposal will also single out position as privileged, because interactions are typically position-dependent. This issue of a privileged basis is not specific for our particles problem, but has a general significance that in particular affects non-collapse interpretations of quantum mechanics (cf.\ \citep{lombardidieks,lombardidieks2}).} The thus defined quantum particles can of course be labeled, on the basis of their individuating physical characteristics. However, these new  labels do not coincide with the factor indices occurring in the original total quantum state---the latter remain evenly distributed over all one-particle states, even in the classical limit. Our proposal is therefore squarely anti-factorist.\footnote{\citet{leegwater} suggest the possibility of constructing a \textit{new} tensor product Hilbert space, once anti-factorist one-particle states of the kind we have discussed have been defined as states of the component particles. Their aim is to tailor this new space in such a way that in it the many-particles state can be written as a product state, with the component particles corresponding to the new factor spaces. However, even if this all works well mathematically, it does not rehabilitate the factorism that we are discussing, namely the doctrine that the factor space labels in the \emph{original} tensor product space refer to particles.} 

\section{Non-fundamentality of the notion of a particle}\label{nonfund}

According to the proposal that we have discussed, states obtained by antisymmetrizing products of one-particle states can be understood as representing particles that possess their own distinctive properties. But obviously, not all antisymmetric states are thus derivable from product states.  Actually, the EPR-Bohm state (\ref{EPReq}) itself is a counterexample: although its spin part  $\{ |\!\uparrow \rangle_1 |\! \downarrow \rangle_2 - |\!\downarrow \rangle_1 |\! \uparrow \rangle_2 \}$ has the form of an antisymmetrized product, the complete state displays a more complicated form of entanglement. This fact is responsible for the non-factorizability of joint probabilities of measurement outcomes on the two wings of the Bell experiment, and consequently for violations of the Bell inequality and for non-locality. By contrast, the ``particles state'' 
\begin{equation}\label{bellvar}
    \frac{1}{\sqrt{2}} \{| L \rangle_1 | R \rangle_2 |\!\uparrow \rangle_1 |\! \downarrow \rangle_2 \ - | R \rangle_1 | L
\rangle_2 |\!\downarrow \rangle_1 |\! \uparrow \rangle_2\}
\end{equation}
does result from antisymmetrizing a product state and leads to probabilities for spin measurements on the two wings of the Bell experiment that do factorize. This means that in the latter state there will be no violations of Bell inequalities and no no-go results for local models. In fact, the particle interpretation that we have outlined immediately provides a local account: the state (\ref{bellvar}) describes a situation in which there are two particles, one on the left and one on the right---the particle on the left-hand side having spin up and the right-hand particle spin down.

The essential difference between the states (\ref{EPReq}) and (\ref{bellvar}) is that in (\ref{bellvar}) a strict correlation exists between spatial and spin states, which is not the case in (\ref{EPReq}). Therefore,  (\ref{bellvar}) represents two particles labeled by ($L, \uparrow$) and ($R, \downarrow$), respectively (note again that these labels differ from the factor labels $1$ and $2$). Both of these particles possess a complete set of one-particle properties. In (\ref{EPReq}) we cannot similarly define co-instantiated sets of particle properties. Therefore, (\ref{bellvar}) lends itself to a straightforward particle interpretation but (\ref{EPReq}) does not.

This example illustrates two things. First, the possibility of a sensible particle interpretation of states of ``identical particles'' is not at all \textit{a priori} given. In fact, most of such states will \emph{not} allow a particle interpretation. This ties in with the second point illustrated by the example, namely that typical and fundamental quantum features like holism and non-locality manifest themselves exactly when a particle picture is \textit{not} appropriate. 

Given the fundamental nature of non-locality and holism in quantum theory, it follows that the notion of a particle cannot be basic in this theory. As we have seen, there do exist quantum states that can consistently be interpreted in terms of constituent particles---in these cases the particle picture bridges the gap between quantum and classical views and helps us to understand the classical limit. But in situations that are fundamentally quantum, as in the Bell experiment and other cases of non-trivial entanglement, particle pictures may mislead rather than clarify. 
Consequently, the particle concept should be considered as \emph{emergent} rather than fundamental: it is precisely when physical mechanisms come into play that wash out the typical quantum effects like non-locality (decoherence plays a pivotal role here) that the notion of a particle becomes applicable and fruitful.

\section{Conclusion}

According to classical physics particles constitute a basic category of what there exists in the physical universe. Like everyday objects, such particles are characterized by distinctive packages of properties. For example, each electron has its own position, in addition to values for mass, charge and momentum. As we have seen, there certainly exist situations in quantum mechanics in which a very similar picture can be used, even in the case of permutation invariant identical particles states. The state given in Eq.(\ref{bellvar}) furnishes a typical example. 

It is important that a distinguishable particle interpretation of such permutation invariant states can be given. Without this possibility there could be no transition, in the classical limit, from quantum to classical particles. Moreover, the way in which the concept of a particle is actually used in experimental practice would not connect with physical theory. But this important possibility of interpreting permutation invariant states in terms of distinguishable particles requires the rejection of factorism. In our opinion this amounts to a refutation of factorism.  

In states that are not (anti)symmetrized products, like the one of Eq.(\ref{EPReq}), physical properties are not bound together in complete packages of one-particle properties. A particle picture  is in such cases not fully adequate\footnote{One may consider the option of thinking in terms of \emph{incomplete} packages of particle properties in such situations. In the case of the state (\ref{EPReq}) this makes it possible to speak about the system on the left and the system on the right, without assigning these systems complete sets of definite particle properties. This strategy accords with the characterization of the situation often found in physical practice, in which one speaks of a left and a right particle sharing a global spin state. Compare the final paragraph of this section.}. 

Such more general, non-fully-particle quantum states are typical of situations that manifest quantum behavior relating to non-locality, holism, and other effects of non-trivial entanglement (entanglement that cannot be reduced to the effect of the (anti)symmetrizing of product states). Experimental evidence that the world fundamentally displays such non-classical behavior---even in cases in which we would \textit{prima facie} expect that classical physics is fully adequate---is accumulating rapidly. The classically motivated notion of a particle does therefore not sit well with the fundamental constitution of the physical world as described by quantum theory and as revealed in high-precision experiments. Rather, the concept of a particle should be seen as \emph{emergent}: as applicable only if conditions relating to semi-classicality (primarily concerning the washing out of effects of non-trivial entanglement) are satisfied.

Awareness of the generally non-particulate nature of the quantum world may provide novel conceptual tools for understanding quantum processes. For example, as the state of Eq.(\ref{EPReq}) shows, it is possible to have quantum states that manifest localization in individual narrow regions in space, without \emph{complete} packages of one-particle properties correlated to these individual regions. The spin part of the EPR-Bohm state does not combine with the spatial part to form a description of two individual particles-with-spin. Abandoning at least part of the usual particle picture in such cases, and adapting our explanatory schemes to what the formalism is suggesting us, may well provide new conceptual instruments, for instance for comprehending quantum information transfer \citep{die8}. 

\section*{Acknowledgment}
We thank Cord Friebe for comments on an earlier version of this article.

\end{document}